\def\simless{\mathbin{\lower 3pt\hbox
     {$\rlap{\raise 5pt\hbox{$\char'074$}}\mathchar"7218$}}}   
\def\simmore{\mathbin{\lower 3pt\hbox
     {$\rlap{\raise 5pt\hbox{$\char'076$}}\mathchar"7218$}}}   

\documentclass{ws-p9-75x6-50}
\def\hide#1{}

\begin{document}


\title{Fast X-ray Oscillations and General Relativity effects in Neutron
star systems}

\author{Mariano M\'{e}ndez}

\address{Astronomical Institute, University of Amsterdam, the
Netherlands 
and
SRON, National Institute for Space Research, 
Utrecht, the Netherlands. E-mail:
M.Mendez@SRON.nl}


\maketitle

\abstracts{Kilohertz quasi-periodic oscillations (kHz QPOs) are
probably caused by matter in Keplerian orbit at some preferred radius
in the accretion disc around a compact star. In a given source, QPO
frequencies can drift by a few hundred Hz following changes of the
inner disc radius; but the disc cannot move closer to the star than the
radius of the innermost stable circular orbit (ISCO) predicted by
general realtivity, hence the kHz QPO frequencies must be limited by
some maximum frequency.}

Long before kHz QPOs were discovered\cite{iauc6319,iauc6320}, it had
been already proposed$^{3-6}$
that evidence of the ISCO around neutron stars could be observed in
flux variability studies of X-ray binaries: For some equations of
state, the neutron star lies within the radius of the ISCO. Clumps of
matter crossing that radius will no longer be rotationally supported
and will fall extremely rapidly onto the neutron star surface;
effectively, the accretion disc is terminated at that radius. There is
a maximum Keplerian frequency around such a neutron star, corresponding
to the minimum possible radius of the inner edge of the disc;
variability of the X-ray flux produced in the disc at frequencies
larger than $\nu_{\rm K} (r_{\rm ISCO})$ should be strongly suppressed.

KHz QPOs models\cite{mlp98a} suggest that the radius of the inner disc
edge decreases as mass accretion rate, $\dot M$, increases. Hence, when
plotted against a quantity that measures $\dot M$, QPO frequency should
increase, but only until the inner disc edge reaches the ISCO; at that
point QPO frequency should remain constant even if the $\dot M$-related
quantity keeps increasing\cite{mlp98a,kfc97}.

This behavior may have been observed\cite{zhang98} with the {\em Rossi
X-ray Timing Explorer (RXTE)}. Figure 1a shows, for the X-ray binary 4U
1820-30, the frequencies of both kHz QPOs vs.\ X-ray intensity, which
is commonly assumed to be a good measure of $\dot M$. Frequencies
increase more or less linearly with intensity up to $\sim$2500 counts
s$^{-1}$, and from then on they remain constant, even as intensity
increases by $\sim$30\,\%.
\begin{figure}[t]
\centerline{
\epsfig{file=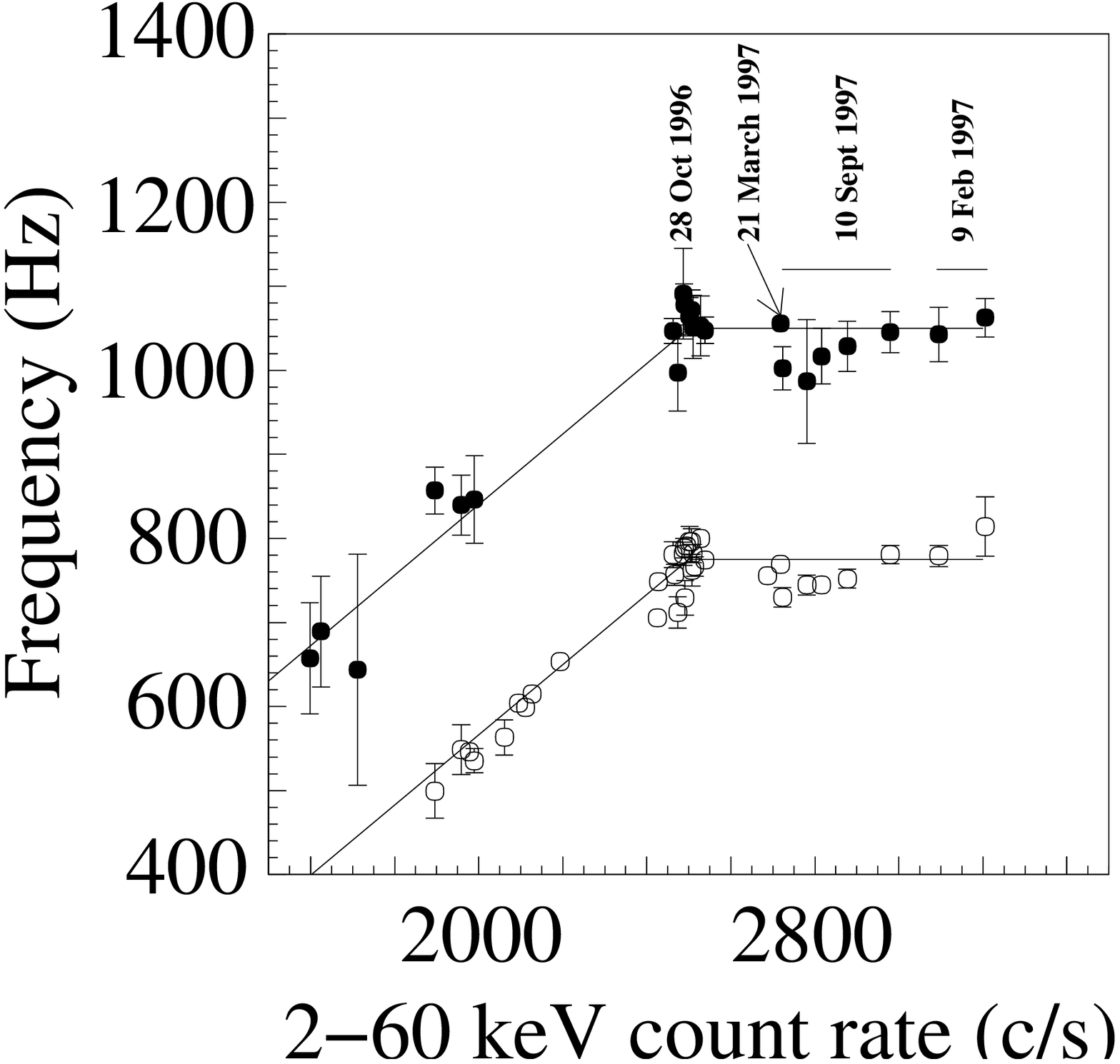, angle=-90, width=4.0cm} 
\epsfig{file=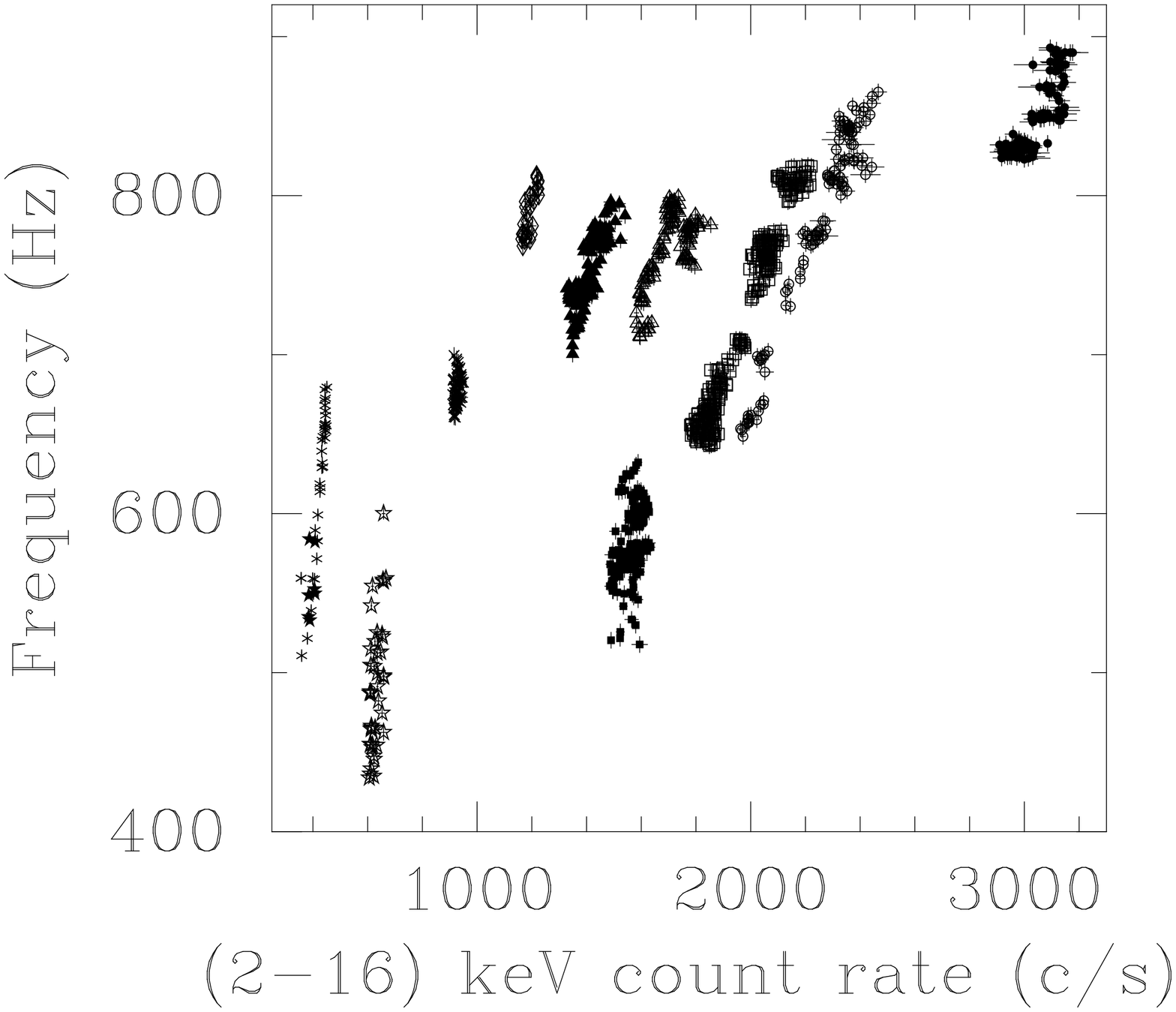, angle=-90, width=4.1cm} 
\epsfig{file=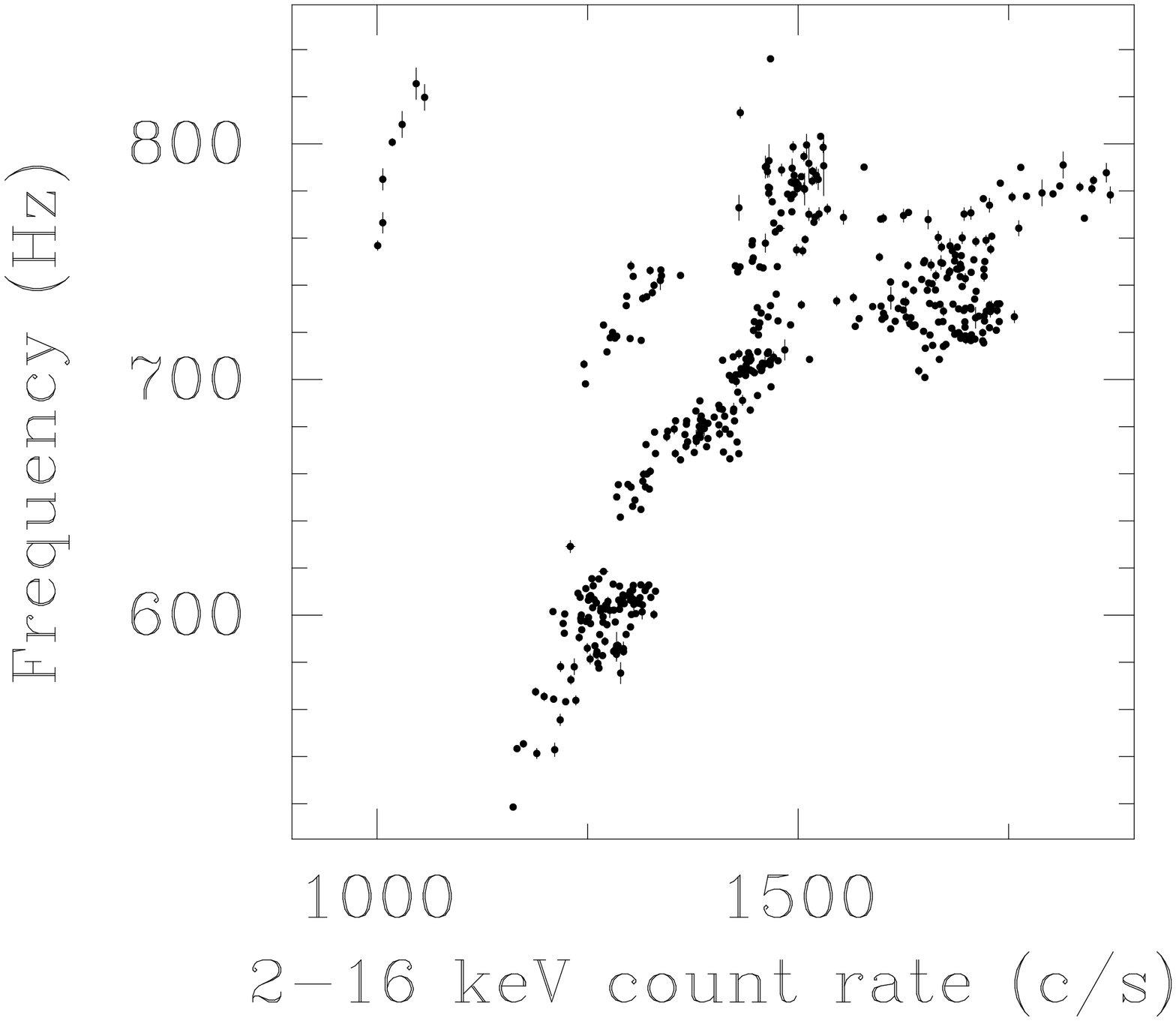, angle=-90, width=4.1cm} 
}
\caption{KHz QPO frequencies vs. X-ray intenisty. {\em Left panel:}
Both kHz QPOs in 4U 1820--30 (data up to June 1998)$^{9}$. {\em Middle
panel:} Lower-frequency kHz QPOs in 4U 1608--52$^{10}$. {\em Right
panel:} Lower-frequency kHz QPOs in 4U 1820--30 (data up to June
2000)$^{11}$}
\end{figure}
It is intriguing that such behavior is not observed in other sources
with kHz QPOs\cite{mendez00}. For instance, Figure 1b shows a similar
plot for one of the kHz QPOs in 4U 1608--52. Interesting in this plot
is the coexsitence of a good frequency-intensity correlation on
timescales shorter than $\sim$1 day (individual segments), with a lack
of correlation on longer timescales. These long-term (uncorrelated)
changes of frequency and intensity seen in 4U 1608--52 and other
sources\cite{mendez00}, could in principle produce a diagram similar to
that shown in Figure 1a. Figure 1c shows QPO frequency vs.\ X-ray
intensity for the lower-frequency kHz QPO in 4U 1820--30, including new
RXTE measurements: It is apparent that, as in 4U 1608--52, there are
long-term uncorrelated variations of frequency and intensity in 4U
1820--30. 

Source intensity may not be a good $\dot M$ tracer, or QPO frequency
may depend upon $\dot M$ through the disc, with disc accretion not
being a fixed fraction of total $\dot M$\cite{kaaret98,mendez99}.
Because in several sources a one-to-one relation between QPO frequency
and spectral properties has been observed\cite{mendez00}, possible
evidence for a QPO frequency saturation vs.\ spectral related
quantities in 4U 1820--30\cite{kaaret99,bloser00} seemed to argue in
favor of the ISCO interpretation for the maximum QPO frequency observed
in this source. However, a careful analysis of the same observations
shows that the evidence of the saturation is not so compelling,
specially when some instrumental corrections, originally not applied,
are taken into account\cite{mendez01}. To fully resolve this issue an
X-ray timing mission with $\sim$10 times the area of
RXTE\cite{barret00} may be needed.

\section*{Acknowledgments} This work was supported by the Netherlands
Organization for Scientific Research, grant PGS 78-277, the Netherlands
Foundation for research in astronomy, grant 781-76-017, the Netherlands
Research School for Astronomy, and LKBF. The author is grateful to
Max-Planck-Institut f\"ur Astrophysik for their hospitality.

\end{document}